# Liquid Crystals with Patterned Molecular Orientation as an Electrolytic Active Medium

**Authors:** Chenhui Peng[1], Yubing Guo[1], Christopher Conklin[2], Jorge Viñals[2], Sergij V. Shiyanovskii[1], Qi-Huo Wei[1] and Oleg D. Lavrentovich[1]*

**Affiliation:**

[1]Liquid Crystal Institute and Chemical Physics Interdisciplinary Program, Kent State University, Kent, OH 44242, USA

[2]School of Physics and Astronomy and Minnesota Supercomputing Institute, University of Minnesota, Minneapolis, MN 55455, USA

Corresponding author:     *olavrent@kent.edu

**Abstract**: Transport of fluids and particles at the microscale is an important theme both in fundamental and applied science. One of the most successful approaches is to use an electric field, which requires the system to carry or induce electric charges. We describe a versatile approach to generate electrokinetic flows by using a liquid crystal (LC) with surface-patterned molecular orientation as an electrolyte. The surface patterning is produced by photo-alignment. In the presence of an electric field, the spatially varying orientation induces space charges that trigger flows of the LC. The active patterned LC electrolyte converts the electric energy into the LC flows and transport of embedded particles of any type (fluid, solid, gaseous) along a predesigned trajectory, posing no limitation on the electric nature (charge, polarizability) of these particles and interfaces. The patterned LC electrolyte exhibits a quadratic field dependence of the flow velocities; it induces persistent vortices of controllable rotation speed and direction that are quintessential for micro- and nanoscale mixing applications.

## 1. Introduction

Electrically driven flows of fluids with respect to solid surfaces (electro-osmosis) and transport of particles in fluids (electrophoresis), collectively called electrokinetics, is a technologically important area of modern science [1,2]. A necessary condition of electrokinetics is separation of electric charges in space. Once separated, these charges are carried by the applied electric field, thus producing electro-osmosis or electrophoresis. The charges might be separated at the solid-fluid interface through dissociation of molecular groups and the formation of electric double layers [3]. The charges can also be separated by the applied electric field, if the solid component is highly polarizable [1,2,4]. Whenever an isotropic fluid is used as an electrolyte, it is the solid component that mediates separation of charges; the fluid is simply supplying counter-ions to complete the double layer build-up.

In this work, we describe a versatile approach to generate electrokinetic effects by using a liquid crystal (LC) with surface-patterned molecular orientation as an electrolyte. The patterned molecular orientation (described by the spatial variations of the so-called director $\hat{\mathbf{n}}$, $\hat{\mathbf{n}}(\mathbf{r}) \neq \text{const}$) is achieved by photo-alignment of the substrates bounding the LC cell. The substrate-imposed pattern imposes director distortions in the entire LC bulk, because of the elastic nature of the LC orientational order. In the presence of a uniform alternating current (AC) electric field, the spatially varying molecular orientation produces space charge that triggers streaming flows of the LC. We demonstrate that the ensuing electrokinetics can transport solid, fluid, and even gaseous inclusions along a predesigned trajectory. The patterned LC electrolyte represents an active matter in which the energy input that drives the system out of equilibrium is localized at the gradients of the orientation order of the medium rather than at the particles embedded in the medium. Since the source of activity is rooted in the properties of the medium, the principle lifts many limitations imposed by isotropic electrolytes on the properties of electrokinetically active interfaces and particles in them (such as the magnitude of surface charge, polarizability, shape asymmetry, etc.).

The principle of the substrate-controlled liquid crystal-enabled electrokinetics (LCEK) can be explained in a greater detail as follows. LCs are anisotropic electrolytes: electric conductivity $\sigma_\parallel$ measured along the average molecular orientation $\hat{\mathbf{n}}$ is usually higher than the conductivity $\sigma_\perp$ measured along the direction perpendicular to it. This anisotropy gives rise to the well-known Carr-Helfrich effect of destabilization of a uniformly aligned LC cell, $\hat{\mathbf{n}}(\mathbf{r}) = \mathrm{const}$ [5,6]. In the electrokinetics approach proposed in our work, the starting point is a LC with a pre-distorted director pattern, $\hat{\mathbf{n}}(\mathbf{r}) \neq \mathrm{const}$, imposed by the LC cell's substrates. The applied electric field $\mathbf{E}$ drags the charges of opposite signs along the curved director lines, accumulating them in different regions of the LC bulk. The gradients of the director thus allow the field to create a non-vanishing volume density of charges $\rho(\mathbf{r})$; the latter depends not only on the conductivity but also on dielectric permittivity of the LC and its anisotropy. The electric field acts on the space charge $\rho(\mathbf{r})$, creating flows of the LC. These flows can carry any particle dispersed in the LC, since the separation of charges occurs in the bulk of the LC medium rather than at (or near) the particle's surface, as in the case of electric double layers around particles in isotropic electrolytes[1,2,4] and charges separated by director distortions near the colloidal particles placed in an otherwise uniform LC [7-12].

An important part of the substrate-controlled LCEK is that the bulk director distortions needed to separate the charges are achieved through patterned photo-alignment. Since the molecular orientation should change from point to point in a pre-described manner, traditional methods of surface alignment such as buffing or more recently proposed rubbing of the substrates with the tip of a cantilever in the atomic force microscopy set up, are not practical. We use a modified version of photo-alignment [13-16], in which the cell substrates are irradiated through plasmonic masks with nanoslits. When such a mask is illuminated with non-polarized light, the slits transmit a polarized optical field that is projected onto a photo-aligning layer. The latter then imposes the desired director field at the substrate and in the adjacent LC.

## 2. Materials and experimental techniques

**2.1 LC materials and cells.** We use a nematic LC with zero dielectric anisotropy ($|\Delta\varepsilon| \leq 10^{-3}$) formed by two components, MLC7026-000 and E7 (both from *EM Industries*) in weight proportion 89.1:10.9. Zero dielectric anisotropy simplifies the experiments and analysis although it is not a necessary requirement to trigger LCEK. The concentration of ions in the mixture is $n_o \approx 10^{19}\,\text{ions}\cdot\text{m}^{-3}$, as measured in the laboratory by a technique described in Ref.[9]. The LC is filled by capillary action between two glass plates with predesigned alignment pattern; both top and bottom plates are treated through the photomask in the assembled state, thus the two photo-induced patterns are the same and establish the distorted director in the LC bulk. The electric field (root-mean square amplitude $E = 40\,\text{mV/µm}$, frequency 5 Hz) is applied in the plane of the cell (along the axis we label as the $x$-axis) by two indium tin oxide (ITO) stripe electrodes, separated by a distance 10 mm . The LC cell is sealed by an epoxy glue, Fig. 1. All experiments were performed at 22°C.

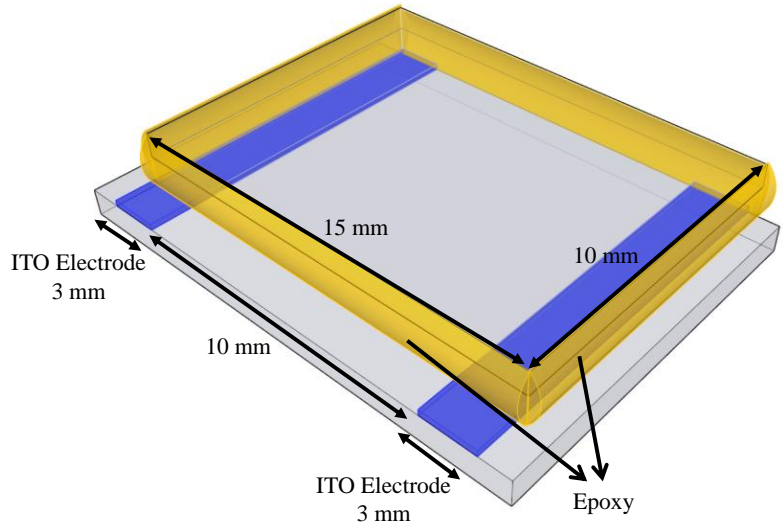

**Figure 1.** **(Color online)** Assembled microfluidic chamber. The chamber is bounded by two glass plates. One of them is a bare glass and the other has stripe indium tin oxide (ITO) electrodes separated by a gap $L = 10$ mm. Both of them are coated with the photoalignment material. The width of each electrode is 3 mm. After the chamber is filled with the liquid crystal, it is sealed by an epoxy.

**2.2 Surface alignment agent.** The photosensitive material Brilliant Yellow (BY) is purchased from *Sigma-Aldrich* and used without further purification. BY is mixed with N,N-Dimethylformamide (DMF) solvent at 1 wt% concentration. In order to improve the stability of BY, reactive mesogen RM257 is mixed with DMF at the concentration 0.2 wt% and then mixed with the solution of BY in DMF (1wt%) in the ratio 1:1 [15,16]. After vortexing for 1 minute, the solution is spin-coated onto two cleaned glass plates. One of them is a bare glass plate and the other one has two patterned indium thin oxide (ITO) electrodes separated by a gap $L = 10$ mm. The glass plates are baked at 95$^{\circ}$C for 30 min. The two glass plates are assembled in a parallel fashion with a gap 50μm between them, set by spherical silica spacers, Fig.1. The fabricated chambers are placed in the optical exposure system and exposed for 30 min. The cell is then filled with the liquid crystal (LC) by capillary action. The exposed BY layer aligns the LC director in the direction parallel to the long side of the slits in the mask. Since the plates are exposed simultaneously, the photoinduced pattern of the director is the same on both surfaces.

**2.3 Optical exposure system for alignment patterning.** The light from the illumination system X-Cite 120 goes through the designed patterns of the photomasks. The masks are made of aluminum films of 150 nm thickness, perforated with rectangular nanoslits, each of a length 220 nm and width 120 nm. The masks are fabricated by the electron beam lithography followed with reactive ion etching. After passing the mask, a nonpolarized light beam becomes polarized. The pattern images are then focused on the microfluidic chamber by the combination of two objective lens. By using one beam splitter, the patterns on the chamber surfaces can be checked by the CCD camera (*Scopetek*). After the director pattern images are focused on the surfaces with photo-alignment materials, Fig. 2, the exposure starts. The exposure time is 30 min. The director is confined to the plane parallel to the bounding plates of the cell. The director patterns are imprinted onto the square region of area 1 mm$^2$; the square is surrounded by the uniformly aligned nematic, $\hat{\mathbf{n}} = (1, 0, 0)$, extending over large distances of 1 cm.

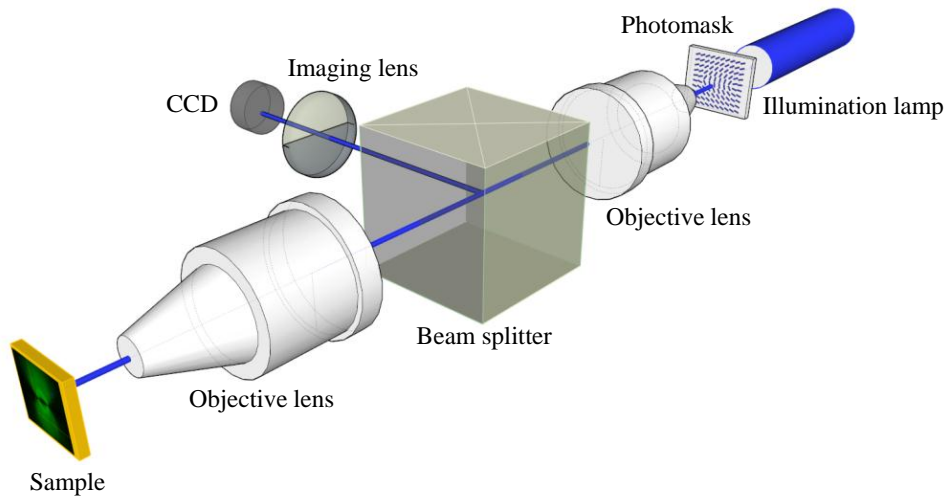

**Figure 2. (Color online)** Photo-patterning optical exposure setup. The light beam from the illumination lamp passes through the predesigned patterns of slits in the photomask and becomes polarized. The pattern image is then focused on the inner surfaces of microfluidic LC chamber by the combination of two objective lenses.

**2.4 Videomicroscopy.** The velocities of electro-osmotic flows are measured under the microscope Nikon Eclipse E600 with a motorized stage (*Prior Scientific*) equipped with a CARV confocal imager (*BD Biosciences, Inc.*) and Photometrics Cascade 650 video camera. The fluorescent illumination system X-Cite 120 is used with the excitation wavelength of 480 nm and emission wavelength of 535 nm. The LC is doped with a small amount ($\sim 0.01\,\text{wt}\,\%$) of tracers, representing fluorescent polystyrene spheres (*Bangs Laboratories*) of diameter $2R = 0.2\,\mu\text{m}$. The small size of the tracers allows one to eliminate the potential influence of dielectrophoretic effects [17]. The microscope was focused at the middle plane of the cell. The tracers cause no visible distortions of the director and are practically non-polarizable. The fluorescent signal of tracers was recorded as a TIFF image with a typical exposure time $\Delta\tau = 325\,\text{ms}$. Using the software package MetaMorph (*Molecular Devices, LLC.*) to superimpose over 1500 images to render a single composite picture, we establish the flow trajectories. The experimental flow velocity fields were obtained using μPIV software PIVlab [18] operated in MATLAB Version R2010b which correlates the position of tracers in consecutive images.

**2.5 PolScope microscopy.** The director fields produced by photo-patterning are established with the help of a polarizing microscope (Nikon E600) equipped with Cambridge Research Abrio LC-PolScope package. The PolScope uses a monochromatic illumination at 546 nm and maps optical retardance and orientation of the optical axis [19].

**2.6 Microfluidic device fabrication for micro-mixing.** A photo-resist Su-8 2025 (MicroChem Corp.) is spin-coated onto the cleaned glass substrates at 500 rpm for 30 s and 1500 rpm for 30 s to create the film with the thickness 50 μm that will be the depth of the channels. Right after photo-resist coating, the substrates are prebaked at $65^\circ\text{C}$ for 2 min and then at $95^\circ\text{C}$ for 8 min. The inlets have the width of 500 μm and the main channel has the width of 1 mm. The angle between the inlets is $40^\circ$, Fig. 10a. The design is exported as a bitmap file and used as a mask in photolithography [20]. After UV exposure for 30 s, the substrates are post-baked at $65^\circ\text{C}$ for 1 min and at $95^\circ\text{C}$ for 5 min. After development in SU-8 developer (MicroChem Corp.) for 5 min, the substrates are rinsed by isopropanol for 1 min in order to form microchannels. The substrate with the microfluidic channel is spin-coated with BY/RM257 mixture at 1500 rpm for 30 s and baked at $95^\circ\text{C}$ for 30 min. Two holes are drilled in the substrate, Fig. 10a, by micro abrasive sand blaster (Problast by Vaniman Manufacturing Co.), in order to provide the inlets for the fluids. This substrate is covered by another glass substrate with patterned ITO electrodes and also coated with the same photoalignment material. This microfluidic chamber is photo-aligned as described above.

**2.7 Micro-mixing efficiency.** To characterize the micro-mixing efficiency, we used an approach based on the so-called standard deviation in the intensity of optical microscopy images of the mixing chamber [21]. The comparison is made for two different modes of mixing, by passive diffusion and by LCEK flows. The time development of mixing is tracked by taking 3000 images by videomicroscopy. Each image contains $N = 653 \times 492 = 321276$ pixels of variable intensity $I_i$, as determined by fluorescent particles. The value of $I_i$ is dimensionless, being normalized by the maximum possible intensity $I_{max}$. In addition to the intensity of each pixel $I_i$, we also calculate the average intensity of each image, $I_{ave} = \frac{1}{N}\sum_{i=1}^{N} I_i$. The standard deviation is defined as $\delta = \sqrt{\frac{1}{N}\sum_{i=1}^{N}(I_i - I_{ave})^2}$. In the unmixed state, the mixing pad is divided into two parts of equal area, one with the maximum fluorescent intensity $I_{max} = 1$ and the other with the minimum intensity $I_{min} = 0$, the average intensity is $I_{ave} = 1/2$, so that $\delta = \delta_0 = \sqrt{\frac{(I_{max} - I_{ave})^2 + (I_{min} - I_{ave})^2}{2}} = 1/2$. In the completely mixed state, the fluorescent intensity $I_i = I_{ave}$, so that $\delta = \sqrt{\frac{1}{N}\sum_{i=1}^{N}(I_{ave} - I_{ave})^2} = 0$.

## 2.8 Numerical simulations of the electro-osmotic flows.

A transport model has been developed to simulate electro-osmotic flows for different photo-patterned arrays, using the Leslie-Ericksen hydrodynamics. The model includes anisotropic mobilities $\mu_{ij} = \mu_\perp \delta_{ij} + \Delta\mu n_i n_j$ where the anisotropic contribution is $\Delta\mu = \mu_\parallel - \mu_\perp$. We consider positive and negative ions of equal concentration $n_o \approx 10^{19}\,\text{ions}\cdot\text{m}^{-3}$. The model is solved numerically using the finite element software package COMSOL in two dimensions, with the parameters of the nematic cell, applied field amplitude and frequency being the same as those used in laboratory experiments, namely, $E = 40\,\text{mV}/\mu\text{m}$ and 5 Hz. Table 1 lists the used parameters. The domain of calculations consists of a square $S_0$ with the side $L_{domain} = 750\,\mu\text{m}$, containing a smaller square $S_1$ with the side length $150\,\mu\text{m}$. The entire domain contains 6294 elements, with $S_1$ more finely meshed than the rest of the domain $S_0$. $S_1$ contains a free triangular mesh with maximum linear size $4.02 \times 10^{-6}\,\text{m}$ and minimum size $1.5 \times 10^{-8}\,\text{m}$. Outside $S_1$, the mesh contains thin quadrilateral boundary layer elements near the edges of $S_0$, with the rest of the domain containing free triangular elements of maximum linear size $1.11 \times 10^{-4}\,\text{m}$ and minimum size $3.75 \times 10^{-7}\,\text{m}$. No slip boundary conditions are imposed on the boundaries of $S_0$ domain. In the resulting simulated electro-osmotic flow patterns, both the charge distribution and the velocity field oscillate in time (as a result of AC driving field), however, the direction of the velocities remains constant. The latter fact reflects the proposed mechanism, according to which the driving force of electro-osmosis represents the product of the field induced charge and the field itself, i.e., the force is proportional to the square of the electric field. Fig. 7c shows the calculated flow for the director pattern presented in Fig. 7a, at the point in which the potential difference across the cell is largest. Four vortices appear around the three disclinations, which are in turn surrounded by eight larger vortices spanning the entire domain of computation (not shown in Fig. 7c). The pattern symmetry and the order of magnitude of the circulated velocity is quite close to that measured experimentally, with a maximum at about 7 µm/s.

**Table 1.** Parameters used in numerical simulations of electro-osmotic flows.

| Parameter | Value | Description |
|---|---|---|
| $d$ | 50 µm | Defect separation |
| $T$ | 295 K | Room Temperature |
| $\mu_\perp$ | $1.2 \times 10^{-19}$ m²/(Vs) | Mobility perpendicular to director |
| $\mu_\parallel$ | $1.7 \times 10^{-19}$ m²/(Vs) | Mobility parallel to director |
| $\varepsilon_\perp$ | 6 | Perpendicular dielectric constant |
| $\varepsilon_\parallel$ | 6 | Parallel dielectric constant |
| $E_0$ | 40 mV/µm | Electric field magnitude |
| $V_0$ | 30 V | Potential difference |
| $f$ | 5 Hz | Frequency of AC field |
| $n_0$ | $10^{19}$ m$^{-3}$ | Concentration of positive and negative ions in domain |

| $K_1$ | $5.3 \times 10^{-12}$ N | Splay elastic constant |
| --- | --- | --- |
| $K_2$ | $2.2 \times 10^{-12}$ N | Twist elastic constant |
| $K_3$ | $7.45 \times 10^{-12}$ N | Bend elastic constant |
| $\alpha_1$ | $6.5 \times 10^{-3}$ Pa s | Leslie-Ericksen viscosity |
| $\alpha_2$ | $-77.5 \times 10^{-3}$ Pa s | Leslie-Ericksen viscosity |
| $\alpha_3$ | $-1.2 \times 10^{-3}$ Pa s | Leslie-Ericksen viscosity |
| $\alpha_4$ | $83.2 \times 10^{-3}$ Pa s | Leslie-Ericksen viscosity |
| $\alpha_5$ | $46.3 \times 10^{-3}$ Pa s | Leslie-Ericksen viscosity |

## 3. Results

**3.1 Spatial charge created by nonuniform director in presence of an electric field.** The necessary condition of electrokinetic motion of the LC is spatial separation of charges. In our approach of the substrate-controlled LCEK, the space charge is induced by the electric field because of the pre-imposed director deformations. Below we derive the space charge density $\rho$ for the director field distorted in the $xy$ plane, $\hat{\mathbf{n}} = \{\cos\alpha(x,y), \sin\alpha(x,y), 0\}$, where $\alpha$ is the angle between $\hat{\mathbf{n}}$ and $x$-axis.

The starting point is the Maxwell's equation for the magnetic field $\mathbf{H}$:

$$\operatorname{curl} \mathbf{H} = \frac{\partial \mathbf{D}}{\partial t} + \mathbf{J}. \qquad (1)$$

Consider a low-frequency harmonic field $\mathbf{E}(t) = \mathbf{E}e^{-i\omega t}$ that creates the current density $\mathbf{J}(t) = \mathbf{J}e^{-i\omega t} = \mathbf{\sigma}\mathbf{E}e^{-i\omega t}$ and the electric displacement $\mathbf{D}(t) = \mathbf{D}e^{-i\omega t} = \varepsilon_0 \mathbf{\varepsilon}\mathbf{E}e^{-i\omega t}$; here $\mathbf{\sigma} = \sigma_\perp \mathbf{I} + \Delta\sigma \hat{\mathbf{n}} \otimes \hat{\mathbf{n}}$ and $\mathbf{\varepsilon} = \varepsilon_\perp \mathbf{I} + \Delta\varepsilon \hat{\mathbf{n}} \otimes \hat{\mathbf{n}}$ are, respectively, the conductivity and dielectric tensors in the laboratory frame, $\Delta\varepsilon = \varepsilon_\parallel - \varepsilon_\perp$ is the dielectric anisotropy, and $\otimes$ is the external product of two vectors, the operation's result is a tensor with components $[\hat{\mathbf{n}} \otimes \hat{\mathbf{n}}]_{ij} = n_i n_j$. We assume that the diagonal components $\sigma_\parallel$, $\sigma_\perp$ of the conductivity tensor and the diagonal components $\varepsilon_\parallel$ and $\varepsilon_\perp$ of the dielectric tensor are frequency independent. Equation (1) can be rewritten as

$$\text{div}\left(\frac{\partial \mathbf{D}}{\partial t} + \mathbf{J}\right) = \text{div}\,\tilde{\mathbf{\sigma}}(\omega)\mathbf{E} = 0, \tag{2}$$

where $\tilde{\mathbf{\sigma}} = \mathbf{\sigma} - i\omega\varepsilon_0 \mathbf{\varepsilon}$ is the effective conductivity tensor.

For low frequency $\omega \ll \sigma_\perp / \varepsilon_0 \varepsilon_\perp \ll c/L$ ($c$ is the speed of light, $L$ is the distance between the electrodes), $\tilde{\mathbf{\sigma}} \approx \mathbf{\sigma}$ and $\mathbf{E} = -\nabla V$, where the potential $V$ obeys the equation $\text{div}(\mathbf{\sigma}\nabla V) = 0$, or

$$\sigma_\perp \nabla^2 V + \Delta\sigma \,\text{div}\left[(\hat{\mathbf{n}} \cdot \nabla V)\hat{\mathbf{n}}\right] = 0, \tag{3}$$

and thus the charge density $\rho = \text{div}\,\mathbf{D}$ reads

$$\rho = -\varepsilon_o \left(\Delta\varepsilon - \varepsilon_\perp \Delta\sigma / \sigma_\perp\right) \text{div}(\hat{\mathbf{n}} \cdot \nabla V)\hat{\mathbf{n}}, \tag{4}$$

Let us consider the external field $E_0$ applied along the *x* axis and assume a weak anisotropy of conductivity, $\Delta\sigma \ll \sigma_\perp$. The electric field acting on the LC can be represented as $\mathbf{E} = \{E_0 + \tilde{E}_x(x,y), \tilde{E}_y(x,y)\}$, where $\tilde{E}_x(x,y)$ and $\tilde{E}_y(x,y)$ are small corrections caused by the director inhomogeneity that satisfy Eq.(4). In the first perturbation order,

$$\sigma_\perp \left( \frac{\partial \tilde{E}_x(x,y)}{\partial x} + \frac{\partial \tilde{E}_y(x,y)}{\partial y} \right) + \Delta\sigma E_0 \left( \cos 2\alpha \frac{\partial \alpha}{\partial y} - \sin 2\alpha \frac{\partial \alpha}{\partial x} \right) = 0, \tag{5}$$

The electric field $\mathbf{E}$ creates the spatially varying charge density

$$\rho = \varepsilon_0 \left( \Delta\varepsilon - \varepsilon_\perp \Delta\sigma / \sigma_\perp \right) E_0 \left( \cos 2\alpha \frac{\partial \alpha}{\partial y} - \sin 2\alpha \frac{\partial \alpha}{\partial x} \right). \tag{6}$$

The field-induced charge density $\rho$ is being acted upon by the applied electric field thus creating a force density $f = \rho E_0 \propto E_0^2$ that causes the flow of the LC controlled by the surface-imposed director pattern. Note that Eq. (6) shows the space charge being dependent on both conductivity anisotropy $\Delta\sigma$ and dielectric anisotropy $\Delta\varepsilon$; either one of them or both can lead to charge separation. In the experimentally studied material, there is no dielectric anisotropy, thus in what follows, we use Eq. (6) with $\Delta\varepsilon = 0$. This choice simplifies the analysis presented in this paper by eliminating the dielectric torques on the director. One should bear in mind, however, that nonzero dielectric anisotropy can be used to create, enhance, or control the space charge, depending on the sign of $\Delta\varepsilon$, as evident from Eq.(6).

**3.2 Electrokinetic flows in LC electrolytes with one-dimensionally periodic director patterns.** Figure 3a, b and c shows three different examples of one-dimensional periodic patterns of the director $\hat{\mathbf{n}} = (n_x, n_y, 0)$ in the photoaligned nematic cell. The patterns are designed as

$$(n_x, n_y) = (|\cos \pi y / l|, \sin \pi y / l) \tag{7a}$$

$$(n_x, n_y) = (\cos \pi y / l, |\sin \pi y / l|) \tag{7b}$$

$$(n_x, n_y) = (\cos \pi (1 - y / l), \sin \pi (1 - y / l)) \tag{7c}$$

in Fig.3 a,b and c, respectively.

In the absence of electric field, the ions are distributed homogeneously in the sample. When the field is applied along the $x$-axis, $\mathbf{E} = (E_0, 0)$, it separates the positively and negatively charged ions along the $y$-axis, by moving them along the "guiding rail" of the director. For example, in Fig. 3c, as easy to see for the shown direction of the field, anisotropic conductivity $\Delta \sigma = \sigma_\parallel - \sigma_\perp > 0$ helps to accumulate positive charges in the regions with the "horizontal" director, $n_x = 1, n_y = 0$, and the negative ones in the regions with $n_x = 0, n_y = 1$. The general expression for the space charge density given by Eq. (6) can be applied to each of the three patterns in Eq.(7) to find the corresponding one-dimensionally varying function $\rho(y)$ (we neglect at the moment edge effects, as the length of stripes is much higher than their width; the issue will be discussed later in this section 3.2). For example, for Eq.(7c) and Fig. 3c, with $\alpha = \pi(1 - y/l)$ ($l$ is the period), according to Eq. (6), in the approximation of low frequency of the field,

$$\rho(y) = \frac{\pi \varepsilon_0 \varepsilon_\perp}{l} \frac{\Delta \sigma}{\sigma_\perp} \cos\left(\frac{2\pi y}{l}\right) E_0. \tag{8}$$

The space charge is steady, as long as there are predesigned director distortions, i.e. the spatial derivatives $\partial \alpha / \partial x$, $\partial \alpha / \partial y$ are nonzero.

Once the charges are separated, the periodic bulk force $f_x(y) = \rho E_0 \propto E_0^2$ acting on the ionic clouds, causes spatially periodic LC electrokinetic flow, Fig. 3d-f. Since the force $f_x$ is proportional to $E_0^2$, the flow does not depend on the field polarity. Reversing polarity of **E** reverses the sign of $\rho$ but the product $\rho$**E** remains unchanged. The feature allows us to use an AC field with zero average to produce the flows. The advantage of the AC driving is that the electrokinetic flows are persistent as long as the field is applied (Supplementary Material Movie S1 [22]); there are no detrimental electrode effects such as field screening and chemical reactions known to occur in the direct current driving case.

The predesigned director pattern in Fig.3a, b, and c controls many essential parameters of the ensuing LCEK flows, such as the viscous resistance to the flow. For example, the pattern in Figs. 3a and d, causes the maximum driving force to be directed along $\hat{\mathbf{n}}$; the flow experiences a relatively low effective viscosity $\eta_\parallel$. In Figs. 3b and e, the flow is mostly perpendicular to $\hat{\mathbf{n}}$, with a higher effective viscosity, $\eta_\perp > \eta_\parallel$, which results in slower velocities. Namely, the maximum amplitudes of velocities are 3.5 μm/s in Fig. 3g, 2.6 μm/s in Fig. 3h, and 2.5 μm/s in Fig. 3i. Note that in the latter case of Fig. 3c, the flows are less regular than in Fig. 3a and b.

It is important to stress here the principal difference between the classic electro-convection phenomena described by the Carr-Helfrich model [5,6] and the electrokinetics described in our work. In the Carr-Helfrich effect [5,6], the LC is aligned uniformly; the director distortions appear as a result of charge separation at director fluctuations; they usually adapt a form of "anomalous" reorientation , linear rolls, two-dimensional array of vortices, determined by the balance of electrohydrodynamic and elastic forces. In our approach, the principal director distortions are predesigned by surface alignment even before the electric field is applied. The concrete shape of these distortions determines the patterns of charge separation and controls the electrokinetic flows when the field is applied.

Furthermore, the patterns in Figs. 3a and b pump an equal amount of LC to the left and right hand sides; the corresponding velocity profiles shown in Fig. 3d and e are symmetric, with zero time average. In Fig. 3g,h, we present the dependencies of the averaged $x$-component of the velocity, $\bar{u}_x(y) = \frac{1}{n}\sum_{i=1}^{n} u_x(x_i, y)$, on the distance measured along the $y$-axis, where $n = 200$ and $u_x(x_i, y)$ is the local velocity component along the $x$-axis, known from the experiment in Figs. 3d and e. The maximum positive and the negative values of the velocity $\bar{u}_x(y)$ are practically the same, Fig. 3g and h. Moreover, the net volumetric flow per unit time, calculated for Figs.3g and h, as $Q_x = \frac{2}{3}h\int_{-y_0}^{y_0} \bar{u}_x(y)dy$ , where $y_0 = 200\,\mu m$ and $h = 50\,\mu m$ is the cell thickness, is practically zero: its deviation from the total volumetric flow defined as $Q_{x,total} = \frac{2}{3} h \int_{-y_0}^{y_0} |\bar{u}_x(y)| dy$ is less than 1%. We conclude that the patterns in Fig.3a,b produce no pumping of the LC along the horizontal axis, as expected from the left-right symmetry of these patterns.

In contrast, the asymmetric pattern shown in Fig.3c, produces a net pumping of the LC to the right hand side, as the overall velocity of flow to the right is higher than that to the left, Fig. 3i, and does not average to zero when integrated over the $y$-coordinate. The volumetric flow $Q_x$ per unit time calculated as above is significant, being about 10% of $Q_{x,total}$. The continuity of the flow is provided by the backflow of the LC from right to left in the homogeneous part of the sample, above and below the patterned area. This can be verified by observing the movement of a LC in a cell in which the patterned region occupies only a fraction of the entire area.

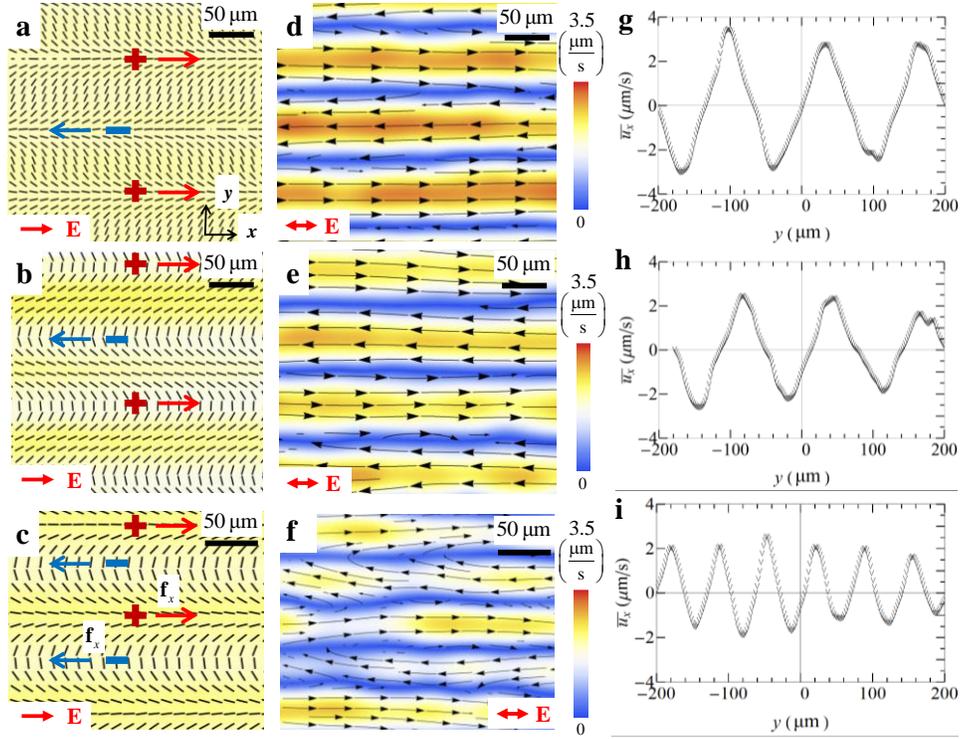

**Figure 3. (Color online)** Electrokinetic flows in LC electrolytes with unidirectionally periodic director patterns. a,b,c, PolScope textures of nematic cell with surface-imposed director patterns described by Eq.(7a, b, c), respectively; "+" and "-" show the charges separated by the electric field of the shown polarity; reversal of field polarity reverses the polarity of charges but the driving Coulomb force remains the same; d, e, f, Corresponding LC flow velocity maps caused by AC field directed along the horizontal $x$-axis; g, h, i, Averaged $x$-component of the nematic velocity vs the distance along the $y$-axis, corresponding to the periodic director fields in parts a, b, and c, respectively.

**3.3 Electrokinetic flows in patterned LC electrolytes with topological defects.** In classic linear electrokinetics, the fluid velocity $\mathbf{u}$ is proportional to the electric field and the resulting flows are irrotational, $\nabla \times \mathbf{u} = 0$. For practical applications such as mixing, it is desirable to trigger flows with vortices [21]. Vortices are very easy to produce in the patterned LC cells, by using localized surface patterns, for example, with topological defects. The topological defects offer yet another degree of freedom in manipulating colloids as they can be used for entrapment and release [23-25].

Director patterns with pairs of disclinations of strength ($m=½$, $-½$) and triplets such as ($m=-½$, $1$, $-½$) and ($m=½$, $-1$, $½$), are created following the general form $n_x = \cos\alpha(x,y)$, $n_y = \sin\alpha(x,y)$, where

$$\alpha(x,y) = m_1 \tan^{-1}\frac{y}{x+d} + m_2 \tan^{-1}\frac{y}{x} + m_3 \tan^{-1}\frac{y}{x-d}, \qquad (9)$$

$d$ is the distance between the cores of two neighboring defects. In all cases, the total topological charge $\sum_i m_i$ is zero, which allows one to smoothly embed the distorted pattern into an otherwise uniform director field. The charge (strength) $m$ is determined by how many times the director rotates by the angle $2\pi$ when one circumnavigates the defect once. In the case of pairs, $m_1 = -m_2 = -1/2$ and $m_3 = 0$. In the case of triplets, $m_1 = m_3 = -1/2, m_2 = 1$. Eq. (9) follows the principle of superposition valid for a director field in a one-constant approximation. These patterns are used to produce the pattern of slits in the photomask and then are reproduced as the true director field in the assembled and photoaligned LC chambers, Figs.4-7.

Figure 4 shows the pattern of slits with a pair of half-integer defects and the main effects it produces in the assembled and photoaligned LC chamber, namely, the distorted director field with two disclinations of half-integer strength, Fig.4c, and the two-vortices electroosmotic flow patterns visualized by fluorescent markers, Fig.4 d. The director and flow patterns are further analyzed in Fig.5.

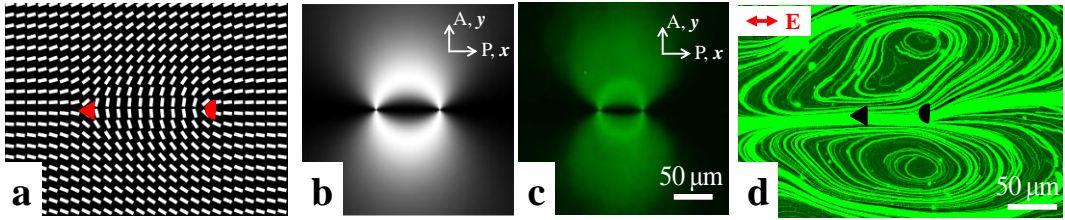

**Figure 4. (Color online)** Disclination pair (-½ , ½) pattern. a, Designed (-½ , ½) director slits pattern; disclination -½ core marked by a triangle and ½ core marked by a semicircle; b, Simulated polarizing microscopy texture of (-½ , ½) disclination pair; c, Polarizing microscopy texture of the (-½ , ½) disclination pattern in the assembled and photoaligned cell; d, Streamlines of electrokinetic flow caused by the AC electric field applied along the $x$-axis and visualized by fluorescent 200 nm tracers. P and A represent polarizer and analyzer.

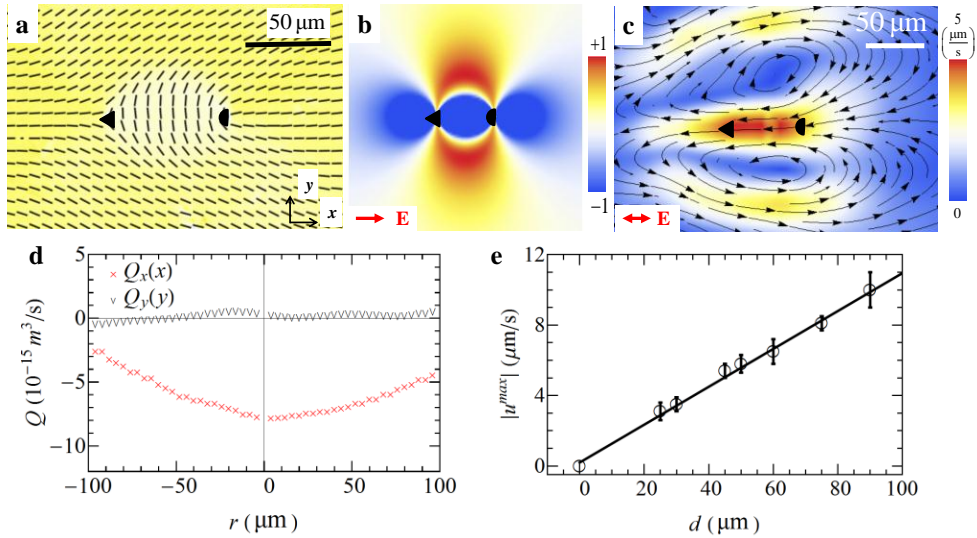

**Figure 5. (Color online)** Nonlinear electrokinetic flows in LC electrolytes with pairs of topological defects. a, PolScope texture of a nematic cell with disclination -½ (core marked by a triangle) and ½ (core marked by a semicircle); b, Map of spatially separated charges in a DC

electric field; c, Velocity of LC flows caused by an AC field $(E_0,0)$; d, LC volume pumped per unit time along the $x$ axis, $Q_x(x)$, and along the $y$ axis, $Q_y(y)$; e, Maximum elecrokinetic flow velocity vs distance $d$ between the disclinations.

Once the electric field is applied, the distorted director, Fig. 5a, creates a local charge density around the disclination pair that is shown in Fig. 5b. The map of positive and negative charges in Fig. 5b is derived from Eq.(6) for $\sigma_\parallel > \sigma_\perp$ and $\varepsilon_\perp = \varepsilon_\parallel$. The resulting electrokinetic flow with two vortices, Fig. 5c, pumps the nematic from ½ defect towards the -½ defect along the line joining the defects and in opposite direction above and below that line.

Pumping efficiency is quantified by the volumetric flow $Q_x(x) = \frac{2}{3}h\int_{-y_0}^{y_0} u_x(x,y)dy$ measured in the vertical $xz$ cross-sections of the cell for each point along the $x$-axis in the range $|x| \leq x_0$, Fig. 5d; $h$ is the cell thickness. Here $x_0 = y_0 = 100\,\mu m$; the horizontal component of velocity $u_x(x,y)$ is known from the experiment, Fig. 5c. The volumetric flow along the $x$-axis, $Q_x(x)$, is negative, reflecting the fore-aft asymmetry of the disclination pair, Fig. 5a. There is no net pumping along the $y$-axis, as $Q_y(y) = \frac{2}{3}h\int_{-x_0}^{x_0} u_y(x,y)dx$ is an antisymmetric function close to zero, Fig. 5d.

The maximum velocity measured in the center of the disclination pair grows linearly with the separation $d$ between the defects, Fig. 5e. This result is understandable since the LCEK velocity $u$ results from the balance of the driving force of density $f_x = \rho E_0 \propto \varepsilon_0 \varepsilon_\perp \Delta\sigma E_0^2 / (d\sigma_\perp)$ and the viscous drag of density $f_{visc} \propto \eta u / d^2$; the estimate is $u \approx \beta \varepsilon_0 \varepsilon_\perp \Delta\sigma d E_0^2 / (\eta \sigma_\perp) \propto d$, where $\beta$ is the numerical coefficient of the order of 1 that depends on the details of director configuration; the latter also defines the actual value of the (generally anisotropic) viscosity $\eta$.

Triplets of defects are produced in a similar way with corresponding patterns of slits, Fig.6. Each triplet is comprised of two half-integer singular disclination at the periphery and a single integer-strength disclination in the center, Fig.6a,e. The integer-strength defects are unstable against splitting into pairs of half-integer disclinations, since the elastic energy of the director field around a disclination of strength $m$ scales as $m^2$ [26]. As a result, the cores of the central defects are seen as split into two closely separated individual cores with two extinction bands; each of these individual cores represent a half-integer disclination, Fig.6c and g. The two cores are kept close to each other by the anchoring forces created by the photopatterned substrates. Because of the finite splitting of the central defect, each triplet configuration is in fact a set of four half-integer disclinations.

In terms of the produced electro-osmotic flows, the main feature of triplets is that they produce four vortices, Fig.6d,h. These patterns are analyzed in a greater detail in Fig.7.

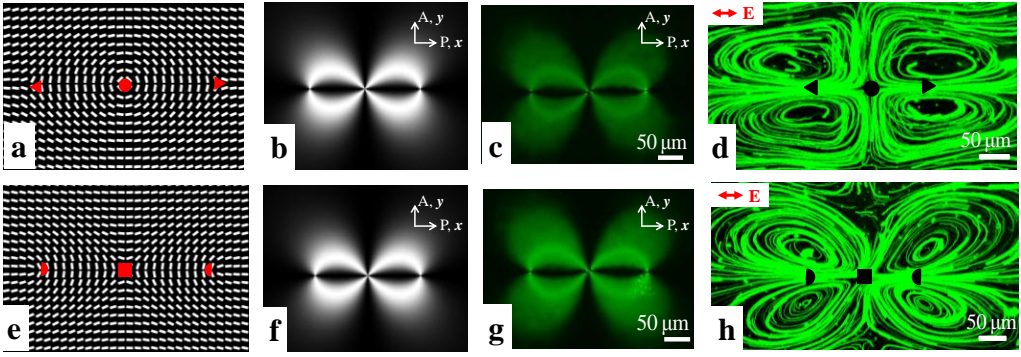

**Figure 6. (Color online)** Slit patterns representing disclination triplets. a and e, Designed slit pattern of (-½, 1, -½) (a) and (½, -1, ½) (e) disclinations; b and f, Simulated polarizing microscopy texture of (-½, 1, -½) and (½, -1, ½) disclinations; c and g, Polarizing microscopy texture of (-½, 1, -½) and (½, -1, ½) sets of disclinations; note a split character of the core of the integer-strength central defect; d and h, Streamlines of electrokinetic flow caused by (-½, 1, -½) and (½, -1, ½) sets, caused by the AC electric field applied along the horizontal direction in the plane of the figure. P and A represent polarizer and analyzer.

The (-½, 1, -½) triplet produces a flow of the "pusher" type, with the fluid moving from the central +1 defect (split into two closely located ½ disclinations) towards the two -½ disclinations at the periphery, Fig. 7b. A complementary triplet (½, -1, ½), produces the "puller" type of flows, with all flow directions being reversed, Fig. 7e. The reason is rooted in the very nature of patterned LCEK, in which the separation of charges and flows depends on the director gradients.

To verify the proposed mechanism of patterned LCEK, we performed numerical simulations of the flows for the three-defects set shown in Fig. 7a. The simulated velocity map, Fig. 7c, is in a good agreement with the experiment, Fig. 7b.

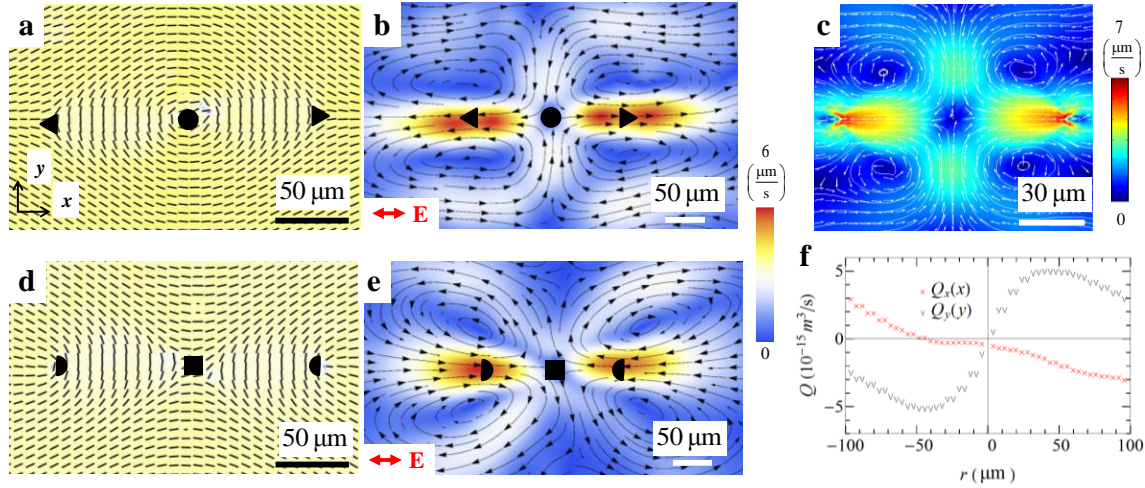

**Figure 7. (Color online)** Nonlinear electrokinetic flows in LC electrolytes with triplets of topological defects. a, PolScope texture of a nematic cell with three disclinations (-½, 1, -½), (-½ core marked by a triangle, 1 core marked by a circle); b, Corresponding velocity map of the nematic caused by the pattern in part (a) caused by an AC electric field $(E_0, 0)$; c, Numerically simulated electrokinetic flow velocity map; d, PolScope texture of disclinations (½, -1, ½), (½ core marked by a semicircle, -1 core marked by a square); e, Corresponding velocity map by the pattern in part (d); f, Volume of nematic fluid pumped per unit time along the horizontal $x$ axis, $Q_x(x)$, and along the $y$ axis, $Q_y(y)$; the flow is of quadrupolar symmetry with four vortices and no pumping effect.

Surface patterning offers a broad freedom in the design of flows. For example, a two-dimensional array of topological defects is designed in the form $n_x = \cos\alpha(x, y)$, $n_y = \sin\alpha(x, y)$, where

$$\alpha(x,y) = \sum_{m,n}\left[\tan^{-1}\frac{y+d_m}{x+d_n} - \frac{1}{2}\left(\tan^{-1}\frac{y+d_m}{x+d_n+d} + \tan^{-1}\frac{y+d_m}{x+d_n-d}\right)\right] +$$

$$\sum_{p,q}\left[\tan^{-1}\frac{y+d_p}{x+d_q} - \frac{1}{2}\left(\tan^{-1}\frac{y+d_p}{x+d_q+d} + \tan^{-1}\frac{y+d_p}{x+d_q-d}\right)\right],$$

$d_m = \sqrt{3}md$, $m = 0, \pm 1, \pm 2...$, $d_n = 3nd$, $n = 0, \pm 1, \pm 2...$, $d_p = \frac{\sqrt{3}}{2}(2p+1)d$, $p = 0, \pm 1, \pm 2...$,

$d_q = \frac{3}{2}(2q+1)d$, $q = 0, \pm 1, \pm 2...$, and $d$ is the distance between the defects of strength 1 and -½.

Typical values of *m*, *n*, *p*, and *q* in the photomasks were 4-5.

The two-dimensional array of vortices of LCEK flows with clockwise and anticlockwise rotation can be easily achieved by this two-dimensional lattice, Fig. 8. Importantly, the polarity of each and every vortex can be reversed by a simple reorientation of the electric field, from $\mathbf{E}=(E_0,0)$, Fig. 8c and d, to $\mathbf{E}=(0,E_0)$, Fig. 8e and f. Another degree of freedom is provided by the reversible character of photoalignment in LCs that can be repeatedly written and re-written [11,12,27,28]. As demonstrated by Hernàndez-Navarro et al [11,12], by using photo-induced trans-cis isomerization that triggers homeotropic-planar realignment at the bounding substrates, one can steer the clusters of LCEK-active colloids with the pear-like shapes. The asymmetric shape creates asymmetric distortions around the particle and enables the propulsion; the role of photo-induced re-alignment is to steer the overall direction of motion. It would be of interest to expand this approach to the substrate-controlled LCEK, by creating and then realigning the director distortions at the substrate that trigger the LCEK flows and transport.

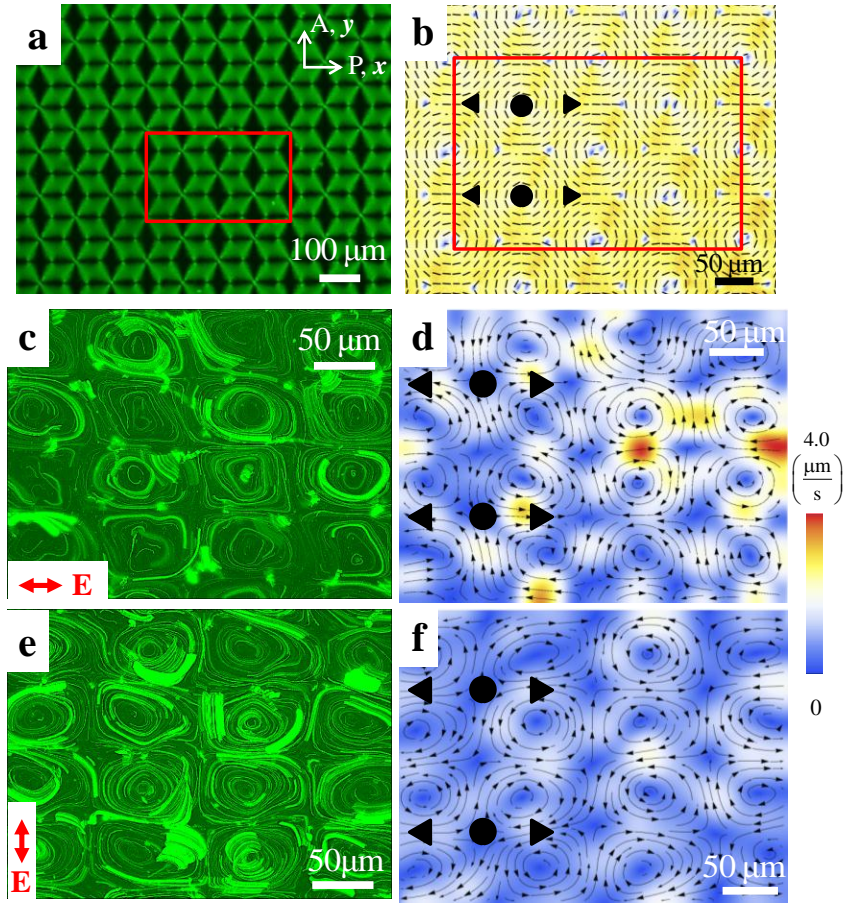

**Figure 8. (Color online)** Nonlinear electrokinetic flows in LC electrolytes with patterns of two-dimensional lattice of topological defects. a, Polarizing microscopy texture of a periodic array of disclinations; b, PolScope texture of the area indicated in part (a); c, Streamlines of electrokinetic flow caused by the AC electric field along the $x$-axis and visualized by fluorescent 200 nm tracers; d, Corresponding velocity field in the same region; e, Streamlines of electrokinetic flow caused by the AC electric field acting along the $y$-axis; f, Corresponding velocity field.

**3.4 Transport of solid, fluid, and gaseous "cargo" in patterned LCEK flows.** Electrokinetic flows can be used to transport particles, such as polystyrene spheres dispersed in the LC, Fig. 9a and b (Supplementary Material Movie S2 [22]), air bubbles (Fig. 9c and d and Supplementary

Material Movie S3 [22]), and droplets of other fluids such as water (Fig. 9e and f and Supplementary Material Movie S4 [22]). The LCEK directed by surface patterning does not impose any limitations on the properties of the "cargo", such as separation of surface charges, polarizability or ability to distort the LC. The latter feature is especially important as compared to the effects of colloidal transport in an otherwise uniform LC cell caused by asymmetric director distortions at the surface of the particle [7-11]. In particular, the polystyrene sphere, Fig. 9a and b, and water droplet, Fig. 9e and f, show tangential anchoring of the director at their surfaces. Because the director distortions in this case are of quadrupolar symmetry, these particles with tangential anchoring do not move in a uniformly aligned LC cell [9]. In the patterned LCEK, however, these particles do move as the electric field energy is rectified at the gradients created in the LC medium by the substrates and there is no need for the transported particle to exhibit any particular surface anchoring properties.

The trajectory of the cargo transport by LCEK is determined by the pattern of molecular orientation. For example, in the conveyor's configurations, the solid sphere, Fig. 9a,b, and air bubble, Fig. 9c,d, move along straight segments. A very different scenario is shown in Fig. 9e,f: the water droplet moves within a vortex along the velocity streams illustrated in Fig.5c and is trapped at the core of the $+1/2$ disclination, joining another water droplets already trapped there; the effect can be used to create micro-scale chemical reactors, described by Sagués et al [10].

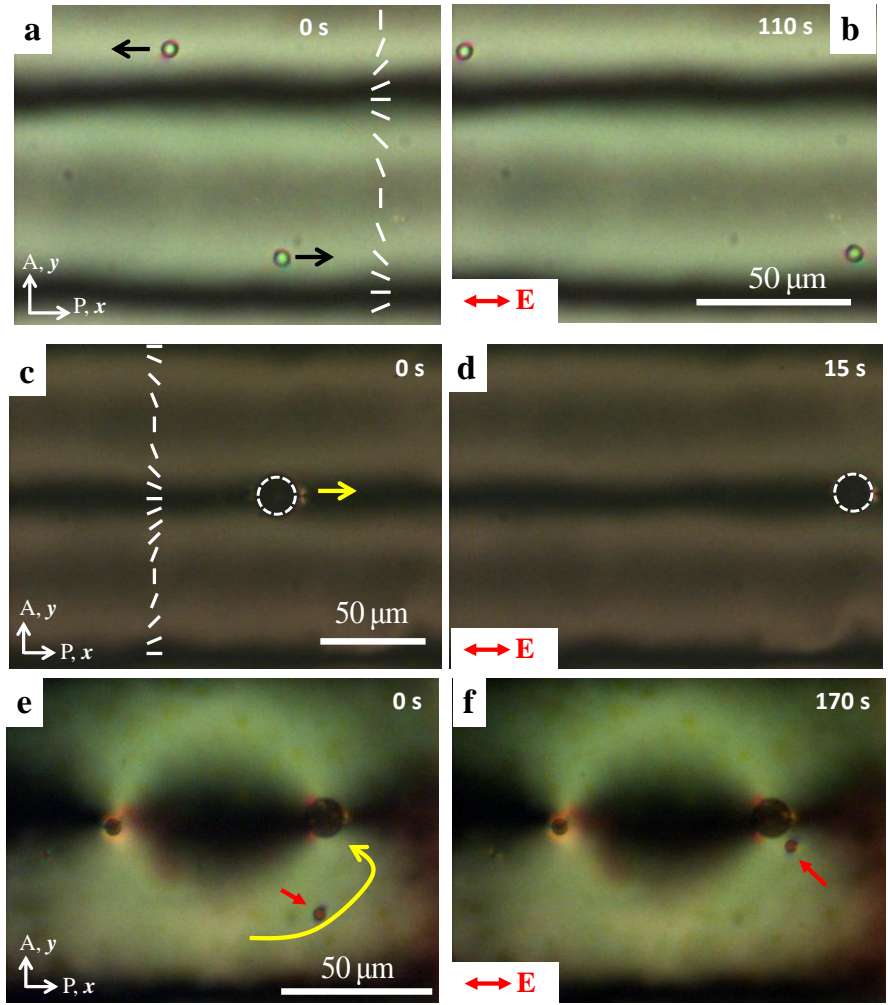

**Figure 9. (Color online)** Transport of cargo by patterned LCEK flows. a and b, Linear transport of two polystyrene particles of diameter 5 μm in the LC with the periodic pattern shown in Fig. 3a. c and d, Linear transport of an air bubble (profiled by a dashed circle) in the nematic chamber with periodic director pattern. e and f, LC flows carry a water droplet (marked by a small arrow) doped with the dye Brilliant Yellow in the nematic chamber with (-½ , ½) disclination pattern shown in Fig. 5a; the trajectory is shown by a curved arrow; the droplet is transported towards the core of the ½ disclination on the right hand side and coalesces with another water droplet that is already trapped there.

**3.5 Micro-mixing by patterned LC-enabled electrokinetic flows.** Surface-imprinted director patterns can be used to facilitate mixing. The circular director distortion is designed as

$$(n_x, n_y) = \left( \cos\left(\tan^{-1}\frac{y}{x}\right)\cos\left(\sqrt{x^2+y^2}\right) - \sin\left(\tan^{-1}\frac{y}{x}\right)\left|\sin\left(\sqrt{x^2+y^2}\right)\right|, \sin\left(\tan^{-1}\frac{y}{x}\right)\cos\left(\sqrt{x^2+y^2}\right) + \cos\left(\tan^{-1}\frac{y}{x}\right)\left|\sin\left(\sqrt{x^2+y^2}\right)\right| \right),$$

Fig. 10b. The cell with these distortions is placed in a microfluidic channel, right after an Y-junction at which two fluids are combined into one channel [29]. The top inlet is injected with a LC containing 200 nm fluorescent tracers and bottom inlet is injected with a pure LC. If there is no electric filed, mixing is achieved only by slow diffusion (low Reynolds number regime). To characterize the mixing efficiency, we plot the normalized standard deviation $\delta/\delta_o$ in fluorescent intensity of the patterns, as the function of time, Fig.10d; here $\delta_o = 1/2$. In the completely unmixed state, $\delta/\delta_o = 1$ and then decreases towards 0 as the two components mix [21,30]. Figure 10d presents a comparison of the mixing efficiency assisted by LCEK (circles) and mixing efficiency of pure diffusion at zero electric field (squares). As clearly seen in Fig. 10d, mixing assisted by LCEK progresses much faster than mixing driven by diffusion only. As compared to other micro-mixers [31,32], the LC mixers do not require any mechanical parts, pressure gradients, nor complicated system of electrodes and ridges obstructing the flow.

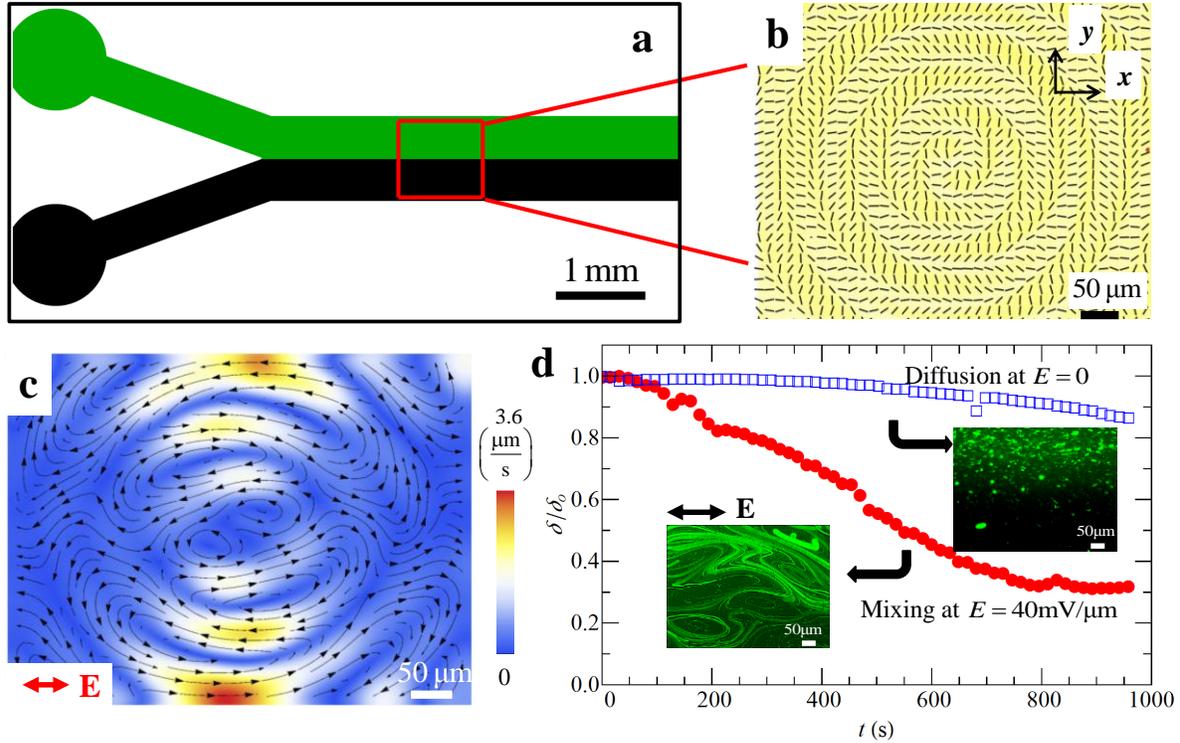

**Figure 10. (Color online)** Micro-mixing in a Y-junction with photo-patterned director distortions. a, Y-junction with a photo-imprinted mixing pad (red square) that combines the LC with fluorescent particles (green) and pure LC (black); b, PolScope texture of the mixing pad; c, Velocity maps within the mixing pad (Supplementary Material Movie S5 [22]); d, Comparison of mixing efficiencies of passive diffusion and LCEK; the insets show the fluorescence microscopy textures of the mixing pad with the exposure time interval 550 s after the start of mixing.

## 4. Discussion and Conclusion

An intriguing question about the experimental setup is how far the director distortions produced by photoalignment at the bounding substrates can propagate into the bulk of the LC. Generally, in absence of any other external aligning factors, the surface-induced alignment is replicated into the LC bulk over macroscopic distances. This is certainly true for the cells used in our experiments, of thickness $50\ \mu m$. In these (and thinner) cells, the disclination lines are joining the top and bottom plates along the vertical $z$-axis, regardless of whether the electric field is applied or not. The situation might change when the thickness of the cell $h$ becomes substantially larger that the characteristic spatial scales $l$ and $d$ of the in-plane director distortions. In this case, the LC might relax through bulk director configurations that are different from the surface patterns. Consider a disclination pair as an example. The elastic energy of a disclination is proportional to its length and to the elastic modulus $K$ of the LC. If $h$ is smaller than the in-plane separation $d$ between the two disclinations, the defects are vertical with the total energy $\sim 2Kh$. If $h > d$, however, the disclinations would tend to reduce their total energy to $\sim 2Kd$ by reconnecting the points at the same substrate [33]. Therefore, the surface-induced pattern of director distortions is expected to persist in the bulk as long as $h \leq l, d$. Furthermore, because of the finite anchoring strength produced by photoalignment, the in-plane director deviates from the imposed surface alignment when the in-plane director gradients become larger than some critical value (equal about $0.2\ \mu m^{-1}$ in our experiments).

To summarize, we demonstrate that the spatially varying director field of an LC electrolyte achieved through photo-imprinted surface alignment allows one to create electrokinetic flows of practically any complexity and vorticity. The flows are persistent, as their velocities are proportional to the square of the applied field, so that the driving field can be of an AC type. The transport of LC and particles dispersed in it is easily controlled by the predesigned director gradients; no mechanical parts and no external pressure gradients are needed. The flow polarity can be changed either by changing the director patterns or the electric field direction. Since the charges are separated in the bulk of electrolytic LC medium rather than at the solid-liquid interfaces, the proposed approach eliminates the need for polarizable/charged interfaces. For example, we demonstrate that LCEK created by surface patterns can carry inclusions such as solid colloids, droplets of water and air bubbles even if these inclusions have no electrophoretic activity (zero charge or zero polarizability) on their own. The cross-sections of the patterned LC microfluidic chambers are not obstructed by any barriers (such as ridges, electrode posts or colloidal particles, needed in other electrokinetic devices), thus combining efficiency of flows with simplicity of design. The approach might find applications in lab-on-the-chip and microfluidic devices of a new type. From the fundamental point of view, the described patterned LC electrolyte represents a new type of active matter in which the energy input that drives the system out of equilibrium occurs locally through orientation distortions of the medium rather than at the particles dispersed in it. This is an essential difference as compared to active materials with artificial or biological swimmers embedded in an otherwise inert surrounding medium such as water [34]. The patterned LC electrolytes add a new dimension to active systems, as now both the medium and the dispersed particles can be used for energy input and departure from equilibrium.

**Acknowledgements**

This work was supported by NSF grants DMR-1507637 (experiment and theory), DMS-1434185 (numerical simulations), and CMMI-1436565 (photopatterning). The authors declare no competing financial interests. We thank Carme Calderer and Dmitry Golovaty for useful discussions.